\documentclass[12pt]{article}
\usepackage{hyperref}
\usepackage{amsmath}
\usepackage{url}
\usepackage{graphicx,color}

\newcommand{\pa}[1]{\left(#1 \right)}

\newcommand{\PPA}[1]{\left(\vbox to 21pt{} #1 \right)}

\def\sinh{{\mathrm{sinh}}}

 \def\ep{{\epsilon}}
 \def\d{{\delta}}
 
 \def\t{{\theta}}

 \def\frac#1#2{{#1\over #2}}
 
 \def\G{{\Gamma}}
 \def\D{{\Delta}}
 \def\g{{\gamma}}
 \def\s{\sqrt}

\def\be{\begin{equation}}
\def\ee{\end{equation}}
\def\ba{\begin{eqnarray}}
\def\ea{\end{eqnarray}}
\numberwithin{equation}{section}

 \def\f {\frac}
 \def\ti{\tilde}

 \def\no{\nonumber \\}

 \def\ep{\epsilon}

\begin{document}

\begin{titlepage}
\thispagestyle{empty}

\begin{flushright}
YITP-17-21
\\
IPMU17-0038
\\
\end{flushright}

\bigskip

\begin{center}
\noindent{{ \large \textbf{Holographic Entanglement Entropy on Generic Time Slices}}}\\
\vspace{2cm}
  Yuya Kusuki$^{a}$, Tadashi Takayanagi$^{a,b}$ and Koji Umemoto$^{a}$
\vspace{1cm}

{\it
$^{a}$Center for Gravitational Physics, Yukawa Institute for Theoretical Physics (YITP),
Kyoto University, Kyoto 606-8502, Japan\\
$^{b}$Kavli Institute for the Physics and Mathematics of the Universe,\\
University of Tokyo, Kashiwa, Chiba 277-8582, Japan\\
}

\vskip 2em
\end{center}

\begin{abstract}
We study the holographic entanglement entropy and mutual information for Lorentz boosted subsystems. In holographic CFTs at zero and finite temperature, we find that the mutual information gets divergent in a universal way when the end points of two subsystems are light-like separated.
In Lifshitz and hyperscaling violating geometries dual to non-relativistic theories, we show that the holographic entanglement entropy is not well-defined for Lorentz boosted subsystems in general. This strongly suggests that in non-relativistic theories, we cannot make a real space factorization of the Hilbert space on a generic time slice except the constant time slice, as opposed to relativistic field theories.

\end{abstract}

\end{titlepage}

\newpage

\section{Introduction}

Entanglement entropy provides us quite a lot of information on quantum states in quantum many-body systems and quantum field theories \cite{Ereview,CCreview,CHreview}. To define the entanglement entropy, we first decompose the total Hilbert space ${\cal H}_{tot}$ into two subsystems ${\cal H}_A$ and ${\cal H}_B$ such that ${\cal H}_{tot}={\cal H}_A\otimes {\cal H}_B$. Then we trace out
the subsystem $B$ and define the reduced density matric $\rho_A$ for the subsystem $A$. The von-Neumann entropy of $\rho_A$ is the entanglement entropy.

\begin{figure}
    \begin{minipage}{0.5\hsize}
      \begin{center}
        \includegraphics[clip, width=60mm]{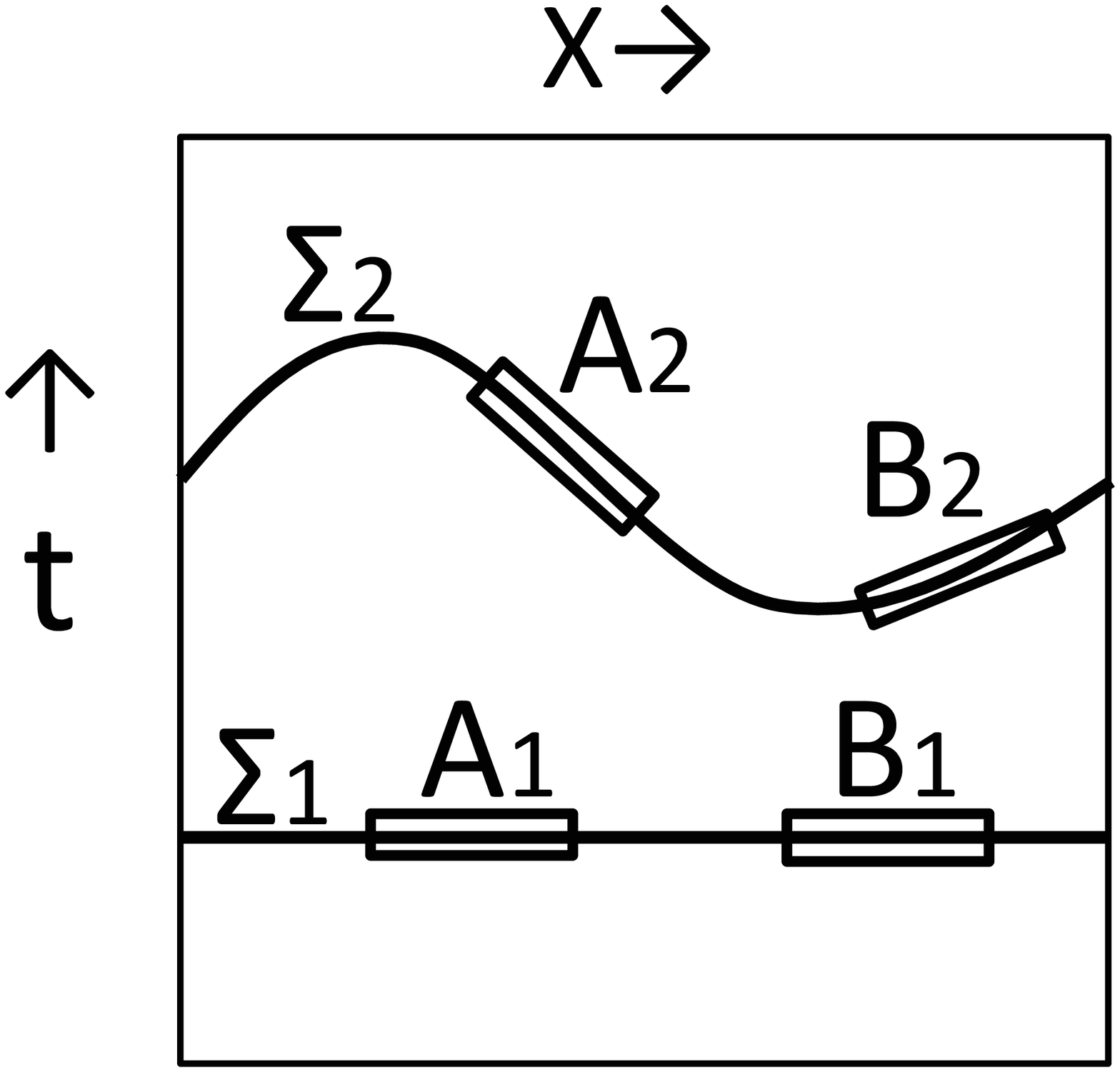}
        \hspace{1.6cm}
      \end{center}
    \end{minipage}
    \begin{minipage}{0.5\hsize}
     \begin{center}
        \includegraphics[clip, width=60mm, angle=270]{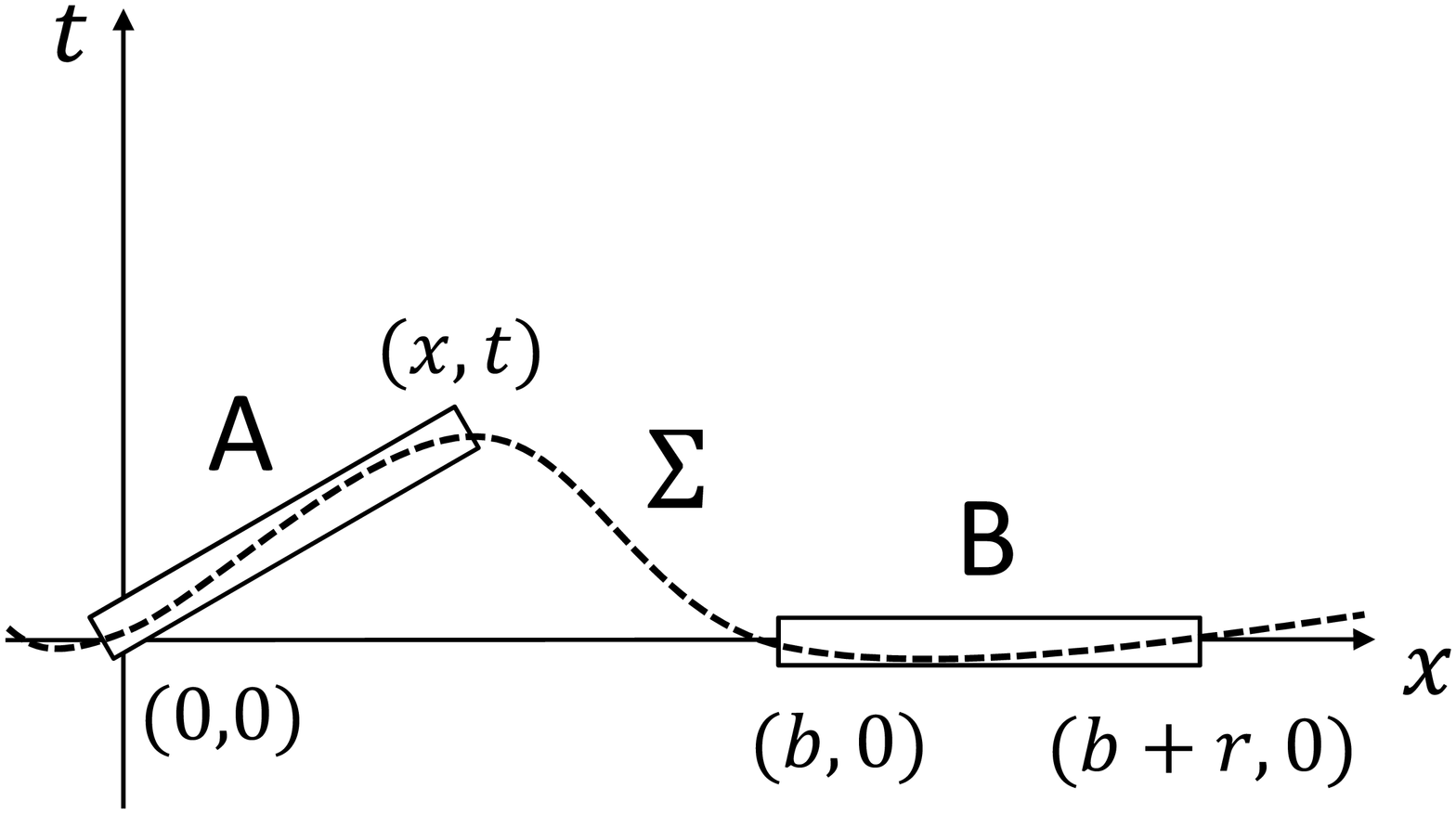}
        \hspace{1.6cm}
      \end{center}
    \end{minipage}
  \caption{The left picture explains entanglement entropy and mutual information can be defined on any space-like time slices in relativistic field theories. A constant time slice is described as $\Sigma_1$, while generic one (deformed one) by $\Sigma_2$. We can define the entanglement entropy $S_{A_1}$, $S_{A_2}$, $S_{B_1}$ and $S_{B_2}$ and their mutual informations. The right picture describes a special setup with a Lorentz boosted interval and an unboosted interval. The dotted curve $\Sigma$ represents the time slice in this setup.}
\label{fig:timeslice}
\end{figure}

In quantum field theories, a quantum state or equally a wave functional is defined on a time slice. Therefore, in order to define entanglement entropy in quantum field theories we need to specify the time slice which defines the quantum state. If we divide the time slice into the region $A$ and $B$, then we have the decomposition of the Hilbert space as ${\cal H}_{tot}={\cal H}_A\otimes {\cal H}_B$. In most of all examples studied so far, we choose the simplest time slice (canonical time slice) defined such that the canonical time $t$ takes a fixed value. However, in Lorentz invariant field theories, we can choose any time slice as long as it is space-like to
define a quantum state. Therefore we can consider entanglement entropy on such generic time slices as in the left picture of Fig.\ref{fig:timeslice}. For example, this plays a very important role in the entropic proof of c-theorem \cite{CaC} and F-theorem \cite{CaF}.

The entanglement entropy and mutual information on generic time slices provide us with more general information of quantum field theories e.g.  properties of ground states and their reduced density matrices. For instance,  in two dimensional conformal field theories (2d CFTs), the mutual information for two intervals on a generic time slice \cite{CaF,BC,CKTW} gives us the information of four point functions for generic values of cross ratios. To study these generalized quantities is the main purpose of this paper. If we consider 2d CFTs, for example, then the simplest entanglement entropy $S_A$ is the one for which $A$ is an interval. If we assume this interval is on a generic time slice, we can always relate it to an interval on a constant time slice by a Lorentz boost and thus the entanglement entropy $S_A$ is essentially reduced to the result for a constant time slice. To get a non-trivial result we consider a mutual information
$I(A,B)=S_A+S_B-S_{A\cup B}$, where $A$ is a boosted interval, while $B$ is not. This is depicted in the right picture of Fig.\ref{fig:timeslice}.  The mutual information for such generalized setups have been studied in two dimensional field theories: refer to \cite{BC} for free field theories and to \cite{CKTW} for orbifold theories (based on the computations in \cite{CCT}).
One of the main aim of this paper is to study this quantity by using the holographic entanglement entropy \cite{RT,HRT,RReview}. In two dimensional CFTs, the special feature of
holographic entanglement entropy and mutual information have been well understood from field theoretic computations \cite{He,Hat,Fn,ABGH}. Our direct calculation using the holographic entanglement entropy allows us to obtain results in higher dimensional CFTs.

On the other hand, in non-relativistic field theories, it is a highly non-trivial question whether we can specify a quantum state by choosing a generic time slice. Therefore the second aim of this paper is to study this question by calculating the holographic entanglement entropy for gravity duals of non-relativistic scale invariant field theories (or so salled Lifshitz-like fixed points) \cite{Lif} and its modification called hyper scaling violating geometries \cite{CGKKM,OTU,HSS}.
As we will see later, if we take a generic time slice, we can define the entanglement 
entropy $S_A$ only when the size of the subsystem $A$ is sufficiently large. This shows that the 
real space factorization of Hilbert space is not always possible on the generic time slices in 
non-relativistic field theories.

This paper is organized as follows. In section two, we analyze the mutual information for boosted
subsystems in AdS$_3/$CFT$_2$ at zero and finite temperature. In section three, we study the mutual information for boosted subsystems in higher dimensional AdS/CFT setups, including a finite temperature case. In section four, we study the holographic entanglement entropy for gravity duals of non-relativistic scale invariant theories when we boost the subsystem. We also analyze the same problem for hyper scaling violating geometries. In section five, we summarize our conclusions.

\section{AdS$_3/$CFT$_2$ Case}

Consider a two dimensional holographic CFT on $R^{1,1}$, whose coordinate is defined by
$(x,t)$. We define the subsystem $A$ and $B$ by two intervals whose end points are
$P_{A,B}$ and $Q_{A,B}$. In particular we choose the points in $R^{1,1}$ as follows (refer to the right picture of Fig.\ref{fig:timeslice}):
\ba
&& P_A=(0,0),\ \ Q_A=(x,t),\no
&& P_B=(b,0),\ \ Q_B=(b+r,0).  \label{itv}
\ea
We are interested in the limit where the interval $A(=[P_A,Q_A])$ and the interval $[Q_A,P_B]$ are both null. Therefore we parameterize
\be
x^+=t+x=2t+\ep_1,\ \ \ x^-=x-t=\ep_1,\ \ \ b=x+t+\ep_2,
\ee
and consider the limit $\ep_1\to 0$ and $\ep_2\to 0$.

The mutual information $I(A,B)$ between $A$ and $B$ is defined as
\be
I(A,B)=S_A+S_B-S_{A\cup B},
\ee
where the UV cut offs cancel out.

The CFT on $R^{1,1}$ is dual to the gravity on Poincare AdS$_3$
\be
ds^2=R^2\left(\frac{dz^2-dt^2+dx^2}{z^2}\right).
\ee
The holographic EE \cite{RT} in AdS$_3/$CFT$_2$ reads
\be
S_A=\frac{L_A}{4G_N},
\ee
where $L_A$ is the geodesic length which connects the two end points of $A$ at the AdS boundary.
The central charge $c$ of the 2d CFT is related to the AdS radius \cite{BrHe}
\be
c=\frac{3R}{2G_N}.
\ee

\subsection{Poincare AdS$_3$}

By applying the AdS$_3/$CFT$_2$, the holographic entanglement entropy $S_A$ is given by
\be
S_A=\frac{c}{6}\log \left[\frac{x^2-t^2}{\delta^2}\right],
\ee
where $\delta$ is the UV cut off or lattice constant. If we set $t=0$, this is reduced to the well-known formula in \cite{HLW,CC}. Note that $S_A$ is well-defined only $|x|>|t|$ i.e. $P_A$ and $Q_A$ are space-like separated.

\begin{figure}
  \centering
  \includegraphics[width=15cm]{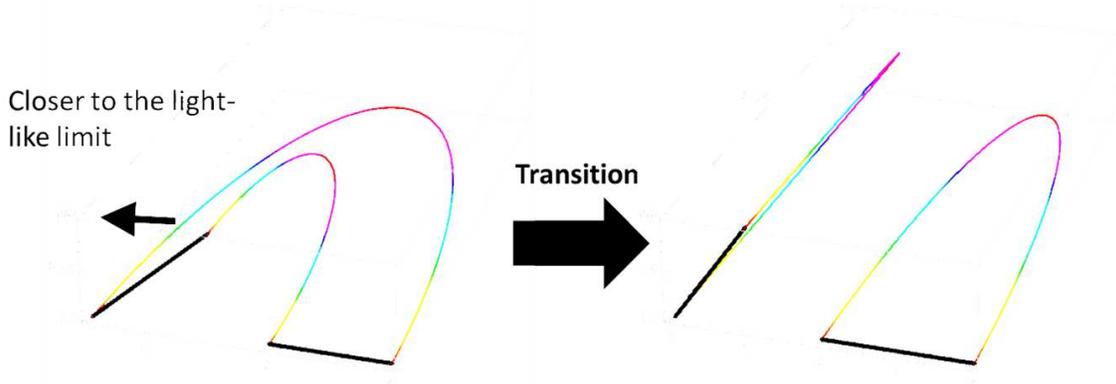}
\caption{The sketch of the transition in the holographic computation of $S_{A \cup B}$.
When the invariant length of subsystem $A$ gets smaller, the extremal surface for the computation of holographic entanglement entropy changes into the disconnected ones (the right picture) and the mutual information becomes vanishing. A similar phase transition occurs when
the invariant length between $Q_A$ and $P_B$ changes.}
\label{fig:trans}
  \end{figure}

In the same way, we can calculate the mutual information $I(A,B)=S_A+S_B-S_{A\cup B}$, assuming $I(A,B)\geq 0$:
\be
I(A,B)=\frac{c}{3}\log\left[\frac{r\s{x^+x^-}}{(b+r)\s{(b-x^+)(b-x^-)}}\right]
\simeq \frac{c}{3}\log\left[\frac{r\s{\ep_1}}{(2t+r)\s{\ep_2}}\right]. \label{mik}
\ee
If the above expression gets negative we should interpret it as  $I(A,B)=0$, which is due to the phase transition phenomena in holographic CFTs \cite{He,Hat,Fn} (refer to Fig.\ref{fig:trans}).

This is comparable to the result in two dimensional rational CFTs (RCFTs) \cite{CKTW}
\be
I(A,B)=\frac{c}{6}\log \frac{2tr}{(2t+r)\ep_2}-\log d_{tot},  \label{xxxr}
\ee
where $d_{tot}$ is the total quantum dimension. In the holographic CFT, we expect
$d_{tot}=\infty$ and this leads to the different result.

The holographic result (\ref{mik}) shows that when $Q_A$ and $P_B$ get close to the light like separation, the mutual information gets divergent as $I(A,B)\sim -(c/6)\log \ep_2$. The same behavior can also be seen in the RCFT result (\ref{xxxr}). On the other hand, the invariant length of subsystem $A$ gets smaller (i.e. $\ep_1$ gets smaller), the mutual information goes to zero. When $\ep_1$ and $\ep_2$ are the same order, the holographic result (\ref{mik}) behaves differently than that for the RCFT (\ref{xxxr}).

\subsection{BTZ Black Hole}

Let us perform the same analysis for the BTZ black hole dual to a finite temperature CFT:
\begin{eqnarray}
ds^{2} & = & \frac{R^2}{z^2}\left(-f(z)dt^{2}+\frac{dz^{2}}{f(z)}+dx^{2}\right),\\
f(z) & = & 1-\frac{z^2}{z^2_H},
\end{eqnarray}
where $R$ is the radius of AdS spacetime and $z_H$ is a positive parameter related to the
inverse temperature via $\beta=2\pi z_H$. The space coordinate $x$ is assumed to be non-compact.

\begin{figure}
  \centering
  \includegraphics[width=8cm]{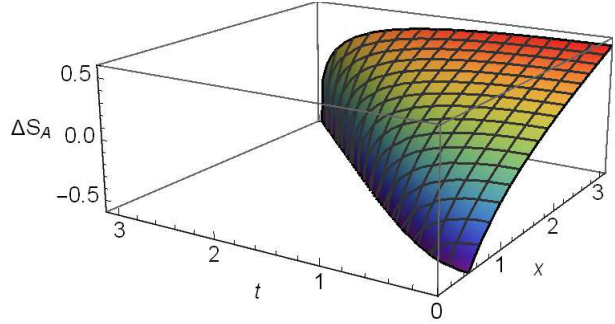}
  \caption{We plotted $S_A$ (\ref{sabtz}) as a function of $x$ and $t$, which are the space and time-like width of the interval $A$ for the BTZ black hole $d=2$. We subtracted the holographic entanglement entropy for a disconnected geodesic from the one for connected one: $\Delta S_{A}=S_{A}-S_{A}^{{\rm (dis)}}$ to remove the UV divergence. Note that the interval $A$ has to be space-like and therefore we need to require $x>t$. We set the parameters $z_{H}=R=G_{N}=1$.}
\label{fig:eeplotBTZ}
  \end{figure}

First, consider a general situation and set the end points as $P_{A}=(0,0)$
and $Q_{A}=(x,t)$. The calculation for HEE can be accomplished
by noting that the BTZ black hole is obtained from a quotient of pure AdS$_3$ as
in \cite{HRT}:
\begin{equation}
S_{A}=\frac{c}{6}\ln\left[\frac{\beta^{2}}{\pi^{2}\delta^{2}}\sinh\left(\frac{\pi}{\beta}(x+t)\right)
\sinh\left(\frac{\pi}{\beta}(x-t)\right)\right]. \label{sabtz}
\end{equation}
Again $S_A$ is well-defined if $P_A$ and $Q_A$ are space-like separated. This is plotted in Fig.\ref{fig:eeplotBTZ}.

Now it is straightforward to compute the mutual information between
arbitrary subregions. In particular we focus on the previous choice (\ref{itv}) and then we obtain
\begin{eqnarray}
I(A,B) & = & \frac{c}{3}\log\left[\frac{\sinh\left(\frac{\pi r}{\beta}\right)}
{\sinh\left(\frac{\pi (b+r)}{\beta}\right)}\right]
+\frac{c}{6}\log\left[\frac{\sinh\left(\frac{\pi}{\beta}(x+t)\right)
\sinh\left(\frac{\pi}{\beta}(x-t)\right)}{\sinh\left(\frac{\pi}{\beta}(x-b+t)\right)
\sinh\left(\frac{\pi}{\beta}(x-b-t)\right)}\right].\no  \label{mutbtz}
\end{eqnarray}

As a check, when we take the zero temperature limit $\beta\to\infty$,
it reproduces (\ref{mik}). In particular, if we use the same limit $\ep_1\to 0$ and $\ep_2\to 0$ in the previous section, we
get the same singularity structure:
\begin{equation}
I(A,B)\simeq \frac{c}{6}\left[\frac{\sinh^2\left(\frac{\pi r}{\beta}\right)\sinh\left(\frac{2\pi t}{\beta}\right)}
{\sinh^2\left(\frac{\pi (b+r)}{\beta}\right)\sinh\left(\frac{\pi b}{\beta}\right)}\right]+\frac{c}{6}\log\left[\frac{\ep_1}{\ep_2}\right].
\end{equation}

\begin{figure}
    \begin{minipage}{0.5\hsize}
      \begin{center}
        \includegraphics[clip, width=60mm, angle=270]{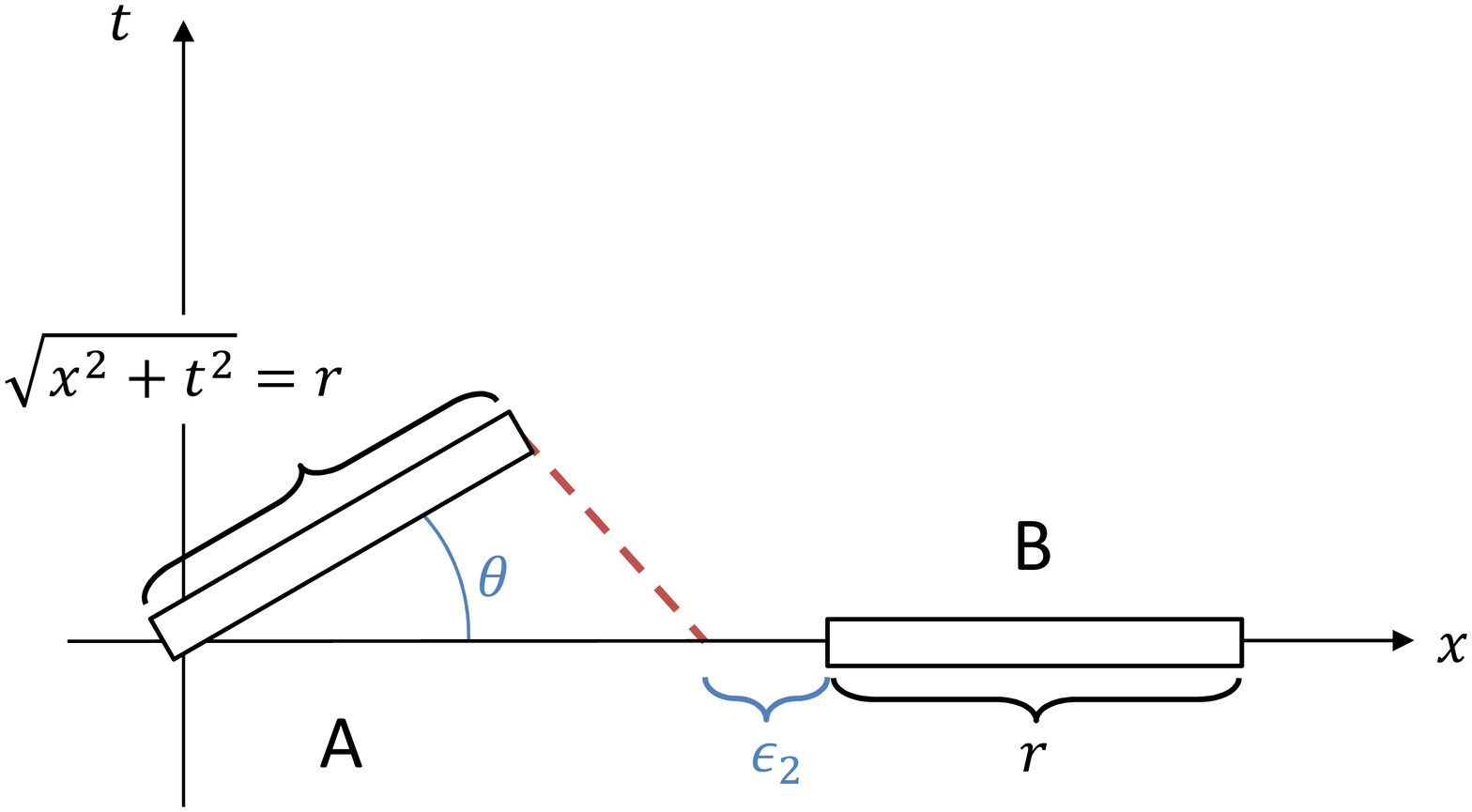}
        \hspace{1.6cm}
      \end{center}
    \end{minipage}
    \begin{minipage}{0.5\hsize}
      \begin{center}
        \includegraphics[clip, width=70mm]{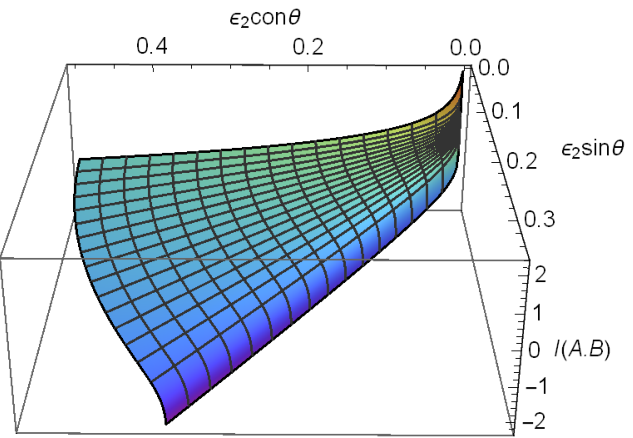}
        \hspace{1.6cm}
      \end{center}
    \end{minipage}
  \caption{In the left graph, we describe the setup of the two intervals parametrized by $\ep_2$ and $\theta$, defined as $(x,t)=(r\cos\theta,r\sin\theta)$ and $b=x+t+\ep_2$. The dashed line denotes a light-like surface. In the right graph, we plotted the mutual information $I(A,B)$ (\ref{mutbtz}) as a function of $(\ep_2,\theta)$ fixing $r=0.5$. The horizontal coordinate and depth coordinate are $\ep_2\cos\theta$ and $\ep_2\sin\theta$.}
\label{fig:miplotBTZ}
  \end{figure}

We focus on the divergence come from the light-like limit between $Q_A$ and $P_B$. The mutual information is plotted in Fig.\ref{fig:miplotBTZ}.

\section{Higher Dimensional Cases}

Here we study the holographic entanglement entropy and mutual information for boosted subsystems in higher dimensional AdS/CFT setups (i.e. AdS$_{d+1}/$CFT$_d$ with $d\geq 3$).

\subsection{AdS$_{d+1}/$CFT$_{d}$}

First, we consider a pure AdS$_{d+1}$ whose metric is given by
\begin{equation}
ds^2=R^2\frac{-dt^2+dz^2+dx^2+\sum_{i=1}^{d-2}dy^2_i}{z^2},
\end{equation}
where $R$ is the radius of anti-de Sitter space. The subsystem $A$ is specified by the boosted strip defined by
\begin{equation}
\label{eq:A}
-\D x/2\leq x\leq \D x/2,~~~-\D t/2\leq t\leq \D t/2, ~~~x=\frac{\D x}{\D t}\cdot t,~~~-\frac{L}{2}\leq y_1,y_2,\ldots,y_{d-2}\leq \frac{L}{2},
\end{equation}
where we take the limit $L\to \infty$. The extremal surface $\g_A$ is specified by the functions $x=x(z)$ and $t=t(z)$. The HEE is computed by extremizing the functional:
\be
S_A=\frac{L^{d-2} R^{d-1}}{2G_N}\int^{z_*}_\delta \frac{dz}{z^{d-1}}\s{1+(x')^2-(t')^2},
\ee
where $\delta$ is the UV cut off  and $L$ is the length of the infinite space length.
The extremal surface extends for the region $\delta\leq z\leq z_*$, where $z=z_*$ is the turning point.

Even though we can find the extremal surface directly, we can easily obtain the final answer by boosting the standard result of holographic entanglement entropy \cite{RT} for the canonical time slice $\D t=0$. If we define the width of the strip by the invariant length $l\equiv \s{(\D x)^2-(\D t)^2}$, then the holographic entanglement entropy is given by the known formula
\begin{equation}
\label{eq:pureAdS}
\begin{aligned}
S_A
&=\frac{L^{d-2}R^{d-1}}{2G_N}\left[ \frac{1}{d-2}\pa{\frac{1}{\d}}^{d-2}-k_d\pa{\frac{1}{l}}^{d-2}\right],\\
\end{aligned}
\end{equation}
where we defined
\be
k_d=\frac{\pi^{(d-1)/2}2^{d-2}}{d-2}
\PPA{\frac{\G\pa{\frac{d}{2(d-1)}}}{\G\pa{\frac{1}{2(d-1)}}}}^{d-1}.
\ee

Then the mutual information $I(A,B)$ for the setup (\ref{itv}) is
\begin{equation}
\label{eq:pureAdSmu}
\begin{aligned}
I(A,B) &= \frac{L^{d-2}R^{d-1}k_d}{4G_N} \left[\left(\frac{1}{r+b}\right)^{d-2}-\left(\frac{1}{\sqrt{x^{2}-t^{2}}}\right)^{d-2}+
\left(\frac{1}{\sqrt{(x-b)^2-t^2}}\right)^{d-2}-\left(\frac{1}{r}\right)^{d-2}\right]\\
&\simeq \frac{L^{d-2}R^{d-1}k_d}{4G_N}\left[-\left(\frac{1}{\sqrt{2t\ep_1}}\right)^{d-2}+
\left(\frac{1}{\sqrt{2t\ep_2}}\right)^{d-2}-\left(\frac{1}{r}\right)^{d-2} +\left(\frac{1}{r+b}\right)^{d-2}\right].
\end{aligned}
\end{equation}
This  shows that when $Q_A$ and $P_B$ get close to the light-like separation, the mutual information gets divergent as $\sim (\ep_2)^{-(d-2)/2}$. On the other hand, as the invariant length of subsystem $A$ gets smaller, the mutual information goes to zero.

\subsection{AdS Black Brane}

We can repeat the computation of $I(A,B)$ for the $d+1$ dimensional AdS black brane (in Poincare coordinate). This metric is given by
\begin{equation}
\begin{aligned}
ds^2&=-R^2 \frac{f(z)}{z^2}dt^2+R^2 \frac{dz^2}{z^2 f(z)}+\frac{R^2}{z^2}dx^2+\frac{R^2}{z^2}\sum_{i=1}^{d-2}dy^2_i,\\
f(z)&=1-\pa{\frac{z}{z_H}}^{d},
\end{aligned}
\end{equation}
where $R$ is the radius of anti-de Sitter space and $z_H$ is the location of the horizon, related to the inverse temperature by $\beta=\frac{4\pi}{d} z_H$. The extremal surface $\g_A$ is specified by the functions $x=x(z)$ and $t=t(z)$. We choose the subsystem $A$ to be defined by (\ref{eq:A}). The holographic entanglement entropy is computed by extremizing the functional:
\be
S_A=\frac{L^{d-2} R^{d-1}}{2G_N}\int^{z_*}_\delta \frac{dz}{z^{d-1}}\s{\frac{1}{f(z)}+(x')^2-f(z)(t')^2},
\ee
where $\delta$ is the UV cut off. The constant $z_*$ describes the turning point where
$t'$ and $x'$ get divergent. The equations of motion for $t$ and $x$ read
\begin{equation}
\begin{aligned}
\frac{f(z)t'}{z^{d-1}\sqrt{\frac{1}{f(z)}-f(z)(t')^{2}+(x')^{2}}} & =  \frac{1}{q},\\
\frac{x'}{z^{d-1}\sqrt{\frac{1}{f(z)}-f(z)(t')^{2}+(x')^{2}}} & =  \frac{1}{p},
\end{aligned}
\end{equation}
where $p$ and $q$ are positive integration constants. These equations can be solved as
\begin{equation}
\begin{aligned}
t' &=\frac{1}{qf(z)\sqrt{\frac{1}{q^{2}}+f(z)\left(\frac{1}{z^{2(d-1)}}-\frac{1}{p^{2}}\right)}},\\
x' &=\frac{1}{p\sqrt{\frac{1}{q^{2}}+f(z)\left(\frac{1}{z^{2(d-1)}}-\frac{1}{p^{2}}\right)}}.
\end{aligned}
\end{equation}
The turning point $z_{*}$ condition allows us to eliminate the parameter $q$ by the relation
\begin{equation}
\label{eq:turning}
\frac{1}{q^{2}}=f(z_{*})\pa{\frac{1}{p^{2}}-\frac{1}{z_{*}^{2(d-1)}}}.
\end{equation}
The sizes of interval $\D t$ and $\D x$ are rewritten as follows:
\begin{equation}
\begin{aligned}
\frac{\D t}{2} & = \frac{1}{q}\int_{0}^{z_{*}}\frac{dz}{f(z)\sqrt{h(z)}},\\
\frac{\D x}{2} & = \frac{1}{p}\int_{0}^{z_{*}}\frac{dz}{\sqrt{h(z)}},
\end{aligned}
\end{equation}
where we introduced the function $h(z)$:
\be
h(z)=f(z)\left(\frac{1}{z^{2(d-1)}}-\frac{1}{p^{2}}\right)
-f(z_{*})\left(\frac{1}{z_{*}^{2(d-1)}}-\frac{1}{p^{2}}\right).
\ee

Now $S_A$ is found as
\begin{equation}
S_{A}=\frac{R^{d-1}L^{d-2}}{2G_{N}}\int_{\d}^{z_{*}}
\frac{dz}{z^{2(d-1)}\s{h(z)}}.
\end{equation}

In order to $z_*$ to be the turning point, $z_*$ should be the smallest solution to
$h(z)=0$. This condition requires
\be
\label{eq:pcon}
\frac{1}{p^2}\leq z_*^{-3d+2}\left(\frac{2(d-1)}{d}-\left(\frac{d-2}{d}\right)z_*^d\right).
\ee
Note that when the inequality is saturated, $z=z_*$ is a double root.\footnote{
At this special value of $z_*$, $\Delta x$ and $\Delta t$ go to infinity due to the double zero. When $z$ is very close to $z_*$, we can find $(\Delta x)^2/(\Delta t)^2\simeq
\frac{q^2(f(z_*))^2}{p^2}=1-\frac{d-2}{2(d-1)} z_*^d \leq 1$. This means that there exist extremal surfaces with $\Delta t>\Delta x$. However such an extremal surface should be discarded when we compute the holographic entanglement entropy as we require the existence of a space-like surface \cite{HHMMR} whose boundary is the union of the subsystem $A$ and its extremal surface $\gamma_A$. Also such a time-like configuration contradicts with the original covariant computation \cite{HRT}.}

\begin{figure}
  \centering
  \includegraphics[width=7cm]{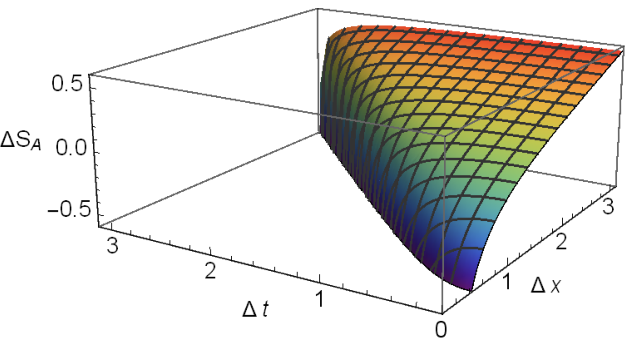}
  \hspace{1cm}
  \includegraphics[width=6cm]{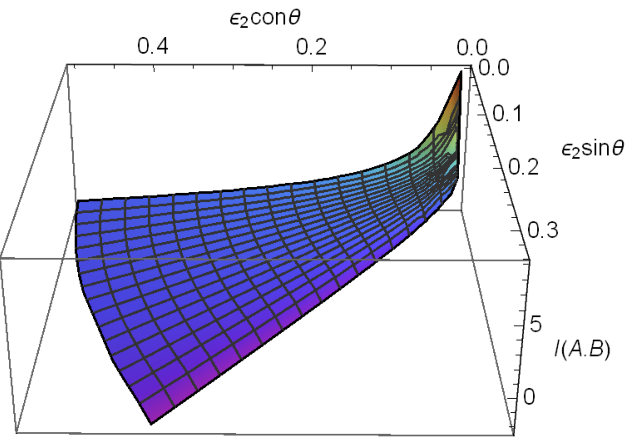}
\caption{In the left graph, we plotted $S_A$ as a function of $\Delta x$ and $\Delta t$, which are the space and time-like width of the strip $A$ for the AdS$_4$ black brane $d=3$. We again subtracted the disconnected entropy from the connected one $\Delta S_{A}=S_{A}-S_{A}^{{\rm (dis)}}$ to remove the UV divergence. Note that the strip $A$ has to be space-like and therefore we need to require $\Delta x>\Delta t$. We set the parameters $z_{H}=R=G_{N}=1$. In the right graph, we plotted the mutual information $I(A,B)$ with the choice of subsystems (\ref{itv}) for the AdS$_4$ black brane as a function of $(\ep_2,\theta)$ defined by $(x,t)=(r\cos\theta,r\sin\theta)$ and $b=x+t+\ep_2$ fixing $r=0.5$, as in the Fig.\ref{fig:miplotBTZ}.}
\label{fig:ee4d}
  \end{figure}

The behaviors of entanglement entropy $S_A$ and the mutual information $I(A,B)$ are plotted in Fig.\ref{fig:ee4d}.

\subsubsection{Light-like Limit}

We are interested in $S_A$ when the subsystem $A$ is boosted such that it is almost null.\footnote{If we take a double scaling limit by combining the zero temperature
 $z_H\to \infty$ limit and the infinite boost, then we get the AdS plane wave geometry \cite{Na}.
 The holographic entanglement entropy was analyzed in \cite{NTT}. Our light-like limit discussed in here is different from this. However we can confirm the same conclusion of the singularity structure also in this double scaling limit.}
This corresponds to the following limit:
\ba
p,q,z_*\to 0,\ \ \mbox{with}\ \  \frac{q}{(z_*)^d}=\mbox{finite},\ \ \frac{p}{(z_*)^d}=\mbox{finite}.  \label{limy}
\ea
In particular we have
\be
\f{q^2}{p^2}=\frac{1}{f(z_*)(1-p^2z_*^{-2(d-1)})}\simeq 1+p^2z_*^{-2(d-1)}, \label{rait}
\ee
where $p^2 z_*^{-2(d-1)}=O(z_*^2)$ is the next leading contribution in the limit.

In this light-like limit, we can approximate $f(z)\simeq 1$ and therefore the computation is reduced to that of the pure AdS. Therefore we get
\begin{equation}
\begin{aligned}
\frac{\Delta t}{2}\simeq \frac{1}{q}\int^{z_*}_{0}\frac{z^{d-1}dz}{\s{1-(z/z_*)^{2(d-1)}}}
&=\frac{\s{\pi}\Gamma\left(\frac{d}{2(d-1)}\right)}{\Gamma\left(\frac{1}{2(d-1)}\right)} \frac{(z_*)^d}{q}, \\
\frac{\Delta x}{2}\simeq \frac{1}{p}\int^{z_*}_{0}\frac{z^{d-1}dz}{\s{1-(z/z_*)^{2(d-1)}}}
&=\frac{\s{\pi}\Gamma\left(\frac{d}{2(d-1)}\right)}{\Gamma\left(\frac{1}{2(d-1)}\right)} \frac{(z_*)^d}{p},
\end{aligned}
\end{equation}
where note that $\Delta t$ and $\Delta x$ are finite in the limit (\ref{limy}). We find
the light-like property $\Delta t\simeq \Delta x$ from (\ref{rait}).

Thus the invariant length of the subsystem $A$ becomes
\be
l=\s{\Delta x^2-\Delta t^2}\simeq 2\frac{\s{\pi}\Gamma\left(\frac{d}{2(d-1)}\right)}{\Gamma\left(\frac{1}{2(d-1)}\right)} z_*.
\ee

Finally the holographic entanglement entropy is estimated as
\ba
S_A&\simeq& \frac{R^{d-1}L^{d-2}}{2G_N}\int^{z_*}_{\delta}\frac{dz}{z^{d-1}\s{1-(z/z_*)^{2(d-1)}}}\no
&=&\frac{R^{d-1}L^{d-2}}{2G_N}\left[\frac{\delta^{-(d-2)}}{d-2}-\frac{\s{\pi}}{d-2}
\frac{\Gamma\left(\frac{d}{2(d-1)}\right)}{\left(\frac{1}{2(d-1)}\right)}z_*^{-(d-2)}\right].
\ea
This reproduces the result in pure AdS (\ref{eq:pureAdS}) as expected.

The mutual information for the choice (\ref{itv}) can be done in the similar way. It is now obvious that in the light-like limit $\ep_1\to 0$ and $\ep_2\to 0$ limit, the dominant singular behavior (i.e. the terms $\sim(\ep_1)^{-(d-2)/2}$ and $\sim(\ep_2)^{-(d-2)/2}$) is exactly the same as
that in pure AdS (\ref{eq:pureAdSmu}), while the finite contributions change.

\begin{figure}
  \centering
  \includegraphics[width=9cm]{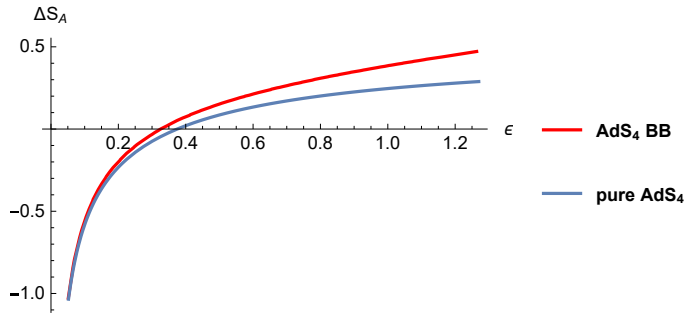}
\caption{We plotted the holographic entanglement entropy $S_A$ for an interval $(\Delta x, \Delta t)=(t+\ep,t)$ as a function of $\ep$ at $t=0.5$. The upper curve describes the AdS$_4$ black brane and the lower one does the pure AdS$_4$. We subtracted the area law divergence $\propto \frac{L}{\delta}$ from the both entropies.}
\label{fig:eeplotep}
  \end{figure}

In the same way, we expect that this singular behavior (\ref{eq:pureAdSmu}) is common to all asymptotically AdS geometries as the only near boundary region is involved in the computation of extremal surface in the light-like limit. In Fig.\ref{fig:eeplotep}, we also numerically confirmed this fact.

\section{Non-relativistic Scale Invariant Theory}

The $d+1$ dimensional gravity background \cite{Lif},
\be
ds^2=R^2\left(-z^{-2\nu}dt^2+z^{-2}dz^2+z^{-2}dx^2+z^{-2}\sum_{i=1}^{d-2}dy^2_i\right),
\ee
is dual to a fixed point (called Lifshitz-like fixed point) with a non-relativistic scale invariance
\be
(t,x,\vec{y})\to (\lambda^\nu t, \lambda x, \lambda \vec{y}).
\ee
The parameter $\nu$ is called the dynamical exponent and the gravity dual is sensible only when
$\nu\geq 1$ \cite{Lif,HK}. Note that $\nu=1$ corresponds to the pure AdS$_{d+1}$.

\subsection{Holographic Entanglement Entropy}

We would like to study the holographic entanglement entropy in this theory.
We choose the strip shape subsystem $A$ which extends in $\vec{y}$ direction
as in (\ref{eq:A}). Its holographic entanglement entropy is computed as the area of extremal surface. The profile of this extremal surface is described by
\be
t=t(z),\ \ \ x=x(z).
\ee
Note that when $t(z)=0$ (the constant time slice), the computation is the same as that for the Poincare AdS$_{d+1}$. Below we would like to focus on more general solutions.

The holographic entanglement entropy $S_A$ is computed by extremizing the functional:
\be
S_A=\frac{L^{d-2} R^{d-1}}{2G_N}\int^{z_*}_\delta \frac{dz}{z^{d-1}}\s{1+(x')^2-z^{2(1-\nu)}(t')^2},
\ee
where $\delta$ is the UV cut off. The constant $z_*$ describes the turning point where
$t'$ and $x'$ get divergent.

The equations of motion for $t$ and $x$ read
\begin{equation}
\begin{aligned}
\frac{z^{2(1-\nu)}t'}{z^{d-1}\s{1+(x')^2-z^{2(1-\nu)}(t')^2}}&=\frac{1}{q},\\
\frac{x'}{z^{d-1}\s{1+(x')^2-z^{2(1-\nu)}(x')^2}}&=\frac{1}{p},
\end{aligned}
\end{equation}
where $p$ and $q$ are integration constants.

The turning point $z_*$ satisfies
\be
1+q^2 z_*^{2(1-\nu)-2(d-1)}-\frac{q^2}{p^2}z_*^{2(1-\nu)}=0.
\ee
Thus we can eliminate $q$ by
\be
\label{eq:qq}
\frac{1}{q^{2}}=z_{*}^{2(1-\nu)}\left(\frac{1}{p^{2}}-\frac{1}{z_{*}^{2(d-1)}}\right).
\ee
The sizes of interval $\Delta t$ and $\Delta x$ are calculated as
\begin{equation}
\begin{aligned}
\frac{\Delta t}{2} & =  \frac{1}{q}\int_{0}^{z_{*}}\frac{dz}{z^{1-\nu}\sqrt{\ti{h}(z)}},\\
\frac{\Delta x}{2}  &=  \frac{1}{p}\int_{0}^{z_{*}}\frac{z^{1-\nu}dz}{\sqrt{\ti{h}(z)}},
\end{aligned}
\end{equation}
where we introduced the function $\ti{h}(z)$:
\begin{equation}
\ti{h}(z)=z^{2(1-\nu)}\pa{\frac{1}{z^{2(d-1)}}-\frac{1}{p^{2}}}-z_{*}^{2(1-\nu)}\pa{\frac{1}{z_{*}^{2(d-1)}}-\frac{1}{p^{2}}}.
\end{equation}
Thus $S_A$ is re-expressed as
\ba
\label{eq:SA}
S_{A} & = & \frac{L^{d-2} R^{d-1}}{2G_{N}}\int_{\delta}^{z_{*}}dz\frac{z^{-2(d-1)+(1-\nu)}}{\sqrt{\ti{h}(z)}}.
\ea

As in (\ref{eq:pcon}), we also have the following condition because $z_*$ should be the smallest solution to  $\ti{h}(z)=0$:
\begin{equation}
\f{1}{p^2}\leq\f{d+\nu-2}{z_*^{2(d-1)}(\nu-1)}.
\end{equation}
And since the left side of (\ref{eq:qq}) should be positive, we can get the other bound for $p$. Hence we finally get the following inequality:
\begin{equation}
\label{eq:domain}
\s{\f{\nu-1}{d+\nu-2}}\leq\f{p}{z_*^{d-1}}\leq1.
\end{equation}

\begin{figure}
  \centering
  \includegraphics[width=9cm]{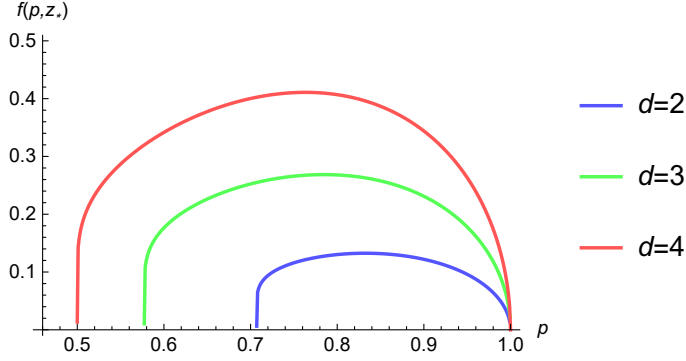}
  \caption{The plot of $f_{\nu,d}(p,z_*)$ as a function of $p$ at $\nu=2$ and $z_*=1$ for
  $d=2$ (blue), $d=3$ (green) and $d=4$ (red). Note that the domain of $f_{\nu,d}(p,z_*)$ perfectly matches (\ref{eq:domain}).}
\label{fig:Lif}
  \end{figure}

It is obvious that there is a (non-relativistic) scale symmetry
\be
(\Delta t,\Delta x, z_*, p)\to (\lambda^{\nu} \Delta t, \lambda \D x,
\lambda z_*, \lambda^{d-1} p).
\ee
Therefore we focus on the ratio
\be
\frac{\Delta t}{(\Delta x)^\nu}=f_{\nu,d}(p,z_*). \label{boundf}
\ee
Actually owing to the scale symmetry, this ratio only depends on the combination:
$z_*^{1-d} p$. Thus we can set $z_*=1$ without loss of generality.
We plotted this function in Fig.\ref{fig:Lif} when $\nu=2$ in various dimensions. As is clear from this plot,
there is an upper bound for this ratio.
\be
\frac{\Delta t}{(\Delta x)^\nu}\leq f^{max}_{\nu,d}.  \label{bound}
\ee
For example, we find
\be
f^{max}_{\nu=2,d=2}=0.132488,\ \ \ f^{max}_{\nu=2,d=3}=0.268549,\ \ \ f^{max}_{\nu=2,d=4}=0.410864.
\ee
Note also that this bound (\ref{bound}) is reduced to the space-like condition $\frac{\Delta t}{\Delta x}<1$ at the Lorentz invariant point $\nu=1$ (i.e. the pure AdS space).

\subsection{Analytical Solution: $d=\nu=2$}

In the special case: $d=\nu=2$, we can analytically find the geodesic \cite{GK}.
First we can solve $z_*$ as
\be
z_*^2=\frac{q^2}{2p^2}-\s{\left(\frac{q^2}{2p^2}\right)^2-q^2}.
\ee
Finally we find
\begin{equation}
\begin{aligned}
\frac{\Delta t}{2}&=\frac{q^2}{4p^2}\log \s{\frac{q+2p^2}{q-2p^2}}-\frac{q}{2}, \\
\frac{\Delta x}{2}&=\frac{q}{2p}\log \s{\frac{q+2p^2}{q-2p^2}}.
\end{aligned}
\end{equation}

The holographic entanglement entropy is found as
\be
S_A=\frac{R}{4G_N}\log \left(\frac{4p^2q}{\delta^2\s{q^2-4p^4}}\right).
\ee

\begin{figure}
  \centering
  \includegraphics[width=6cm]{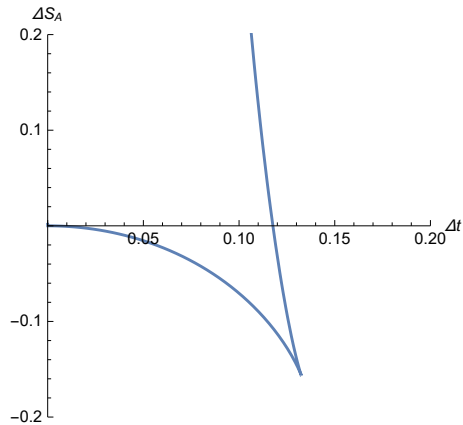}
  \caption{The plot of holographic entanglement entropy $S_A$ as a function of $\Delta t$ at $\Delta x=1$. We subtracted the entanglement entropy for $\Delta t=0$ from $S_A$.
  We set $R=4G_N$ in (\ref{heelift}). There are two branches which correspond to two extremal surfaces. We need to choose the lower blanch for the computation of $S_A$, while the upper one is unphysical.}
\label{fig:Lifhee}
  \end{figure}

If we define $\eta=\frac{2p^2}{q}$, then we find
\be
\frac{\Delta t}{(\Delta x)^2}=\frac{\log\left(e^{-\eta}
\s{\frac{1+\eta}{1-\eta}}\right)}{2\left[\log\left(
\s{\frac{1+\eta}{1-\eta}}\right)\right]^2},
\ee
which has the upper bound mentioned before.
Finally we obtain
\be
S_A=\frac{R}{4G_N} \log \left[\frac{(\Delta x)^2 \eta^2}{\delta^2\s{1-\eta^2}
 \left[\log\left(
\s{\frac{1+\eta}{1-\eta}}\right)\right]^2}\right].  \label{heelift}
\ee
The special value $\eta=0$ corresponds to $\Delta t=0$ and in this case we reproduced the known result $S_A=\frac{R}{2G_N}\log\frac{\Delta x}{\delta}$ \cite{ALT}, which actually takes the identical form as the one in a relativistic CFT. To see the general behavior,  we plotted $S_A$ in Fig.\ref{fig:Lifhee} as a function of $\Delta t$ when we fix $\Delta x=1$. It is monotonically decreasing as $\Delta t$ gets larger, though it is always positive.

\subsection{Consistent Time Slices in Lifshitz Theory}

Now we would like to come back to our original question: can we take generic time slices in non-relativistic scale invariant theories as in relativistic field theories ?
Our analysis shows that the holographic entanglement entropy is defined only when
the condition (\ref{bound}) is satisfied. Consider a deformation of the constant time slice (canonical time slice) $t=0$ to a generic time slice described by $t=t(x)$. If we can define a quantum state on this time slice, we expect that the total Hilbert space ${\cal H}_{tot}$ can be decomposed into the direct product of ${\cal H}_x$ which corresponds to each lattice point $x$ such that
\be
{\cal H}_{tot}=\otimes_{x}{\cal H}_x, \label{dech}
\ee
assuming an appropriate lattice regularization. Then we should be able to define the entanglement entropy for any intervals on this non-canonical time slice. However, by choosing the length of the interval $A$ enough small, we can easily find an interval $A$ on this non-canonical time slice which does not satisfy the bound  (\ref{bound}). This shows that the entanglement entropy is not well-defined\footnote{One might worry that in such cases, the holographic entanglement entropy formula itself can break down. Even though we do not have a solid argument which denies this possibility, it is natural to believe we can still apply the extremal area formula, remembering that it can be derived from the bulk-boundary principle without using special properties of AdS spaces \cite{LM,DLR}. Also note that the Bekenstein-Hawking formula of black hole entropy can be applied to various spacetimes, not only AdS but also flat spaces.}
and shows an inconsistency of the real space decomposition of the Hilbert space (\ref{dech}). This also suggests that a quantum state in the non-relativistic scale invariant theory may be not well described or may be at least highly non-local on a time slice unless it is a constant time slice $t=$const.\footnote{
It is intriguing to note that another interesting and exotic aspect of holographic entanglement entropy, related to entanglement wedge, in the Lifshitz background has been found in \cite{GK}.}
It might be useful to recall that extremal surfaces for holographic entanglement entropy exist only for special subsystems in the NS5-brane gravity backgrounds \cite{RT}, where the dual theory is expected to be non-local.

Let us comment on the mutual information for Lorentz boosted subsystems. If the separation of $Q_A$ and $P_B$ in (\ref{itv}) is almost light-like, the bound (\ref{bound}) is satisfied only if $b-x\simeq
t$ is large enough. In that case, there is no singular behavior in the light-like limit
as we can also see from the plot Fig.\ref{fig:Lifhee}. This is in contrast with the relativistic results in previous sections.

\subsection{Non-relativistic Theory with Hyperscaling Violation}

Finally we would like to study a more general class of gravity duals, called the hyper scaling violating geometry.
\cite{CGKKM,OTU,HSS}. This describes a non-relativistic theory with a scale symmetry violation and its gravity dual is given by the following class of $d+1$ dimensional metric:
\ba
ds^2=R^2\left(z^{-2a}dz^2-z^{-2b}dt^2+z^{-2}dx^2+z^{-2}\sum_{i=1}^{d-2}dy_i^2\right), \label{hsv}
\ea
where $a$ and $b$ are given by
\ba
a=1-\f{\theta}{d-1-\theta},\ \ \ b=1+\frac{(d-1)(\nu-1)}{d-1-\theta}.
\ea
The parameter $\nu$ is the dynamical exponent as before and the new one $\theta$ is the hyper scaling violation
exponent. Obviously the metric (\ref{hsv}) does not have any scale symmetry for $\theta\neq 0$ as opposed to the pervious Lifshitz case (i.e. $\theta=0$).

We can repeat the previous analysis of holographic entanglement entropy $S_A$ for the same choice of strip subsystem $A$. Introducing positive integration constants $p$ and $q$ as before, we find
\ba
&&\f{\Delta t}{2}=\int^{z_*}_0 dz\frac{z^{-a+1}}{{\s{z^{-2b+2}-\frac{q^2}{p^2}z^{-4b+4}+q^2z^{-2(d-1)-4b+4}}}},
\label{dt}\\
&& \f{\Delta x}{2}=\frac{q}{p}\int^{z_*}_0 dz\frac{z^{-a+1-2b+2}}{{\s{z^{-2b+2}-\frac{q^2}{p^2}z^{-4b+4}+q^2z^{-2(d-1)-4b+4}}}}, \label{dx} \\
&& S_A=\frac{L^{d-2} R^{d-1}}{2G_N}\int^{z_*}_{\delta} dz
\frac{q z^{-2d-2b-a+5}}{{\s{z^{-2b+2}-\frac{q^2}{p^2}z^{-4b+4}+q^2z^{-2(d-1)-4b+4}}}}.
\ea
We can eliminate $q$ dependence by $q=\pa{z_*^{1-b}\s{\f{1}{p^2}-\f{1}{z_*^{2(d-1)}}}}^{-1}$.

Actually, even though the full gravity dual does not have any scale symmetry, we can find the following scale symmetry for the integral equations (\ref{dt}) and (\ref{dx}):
\be
(\Delta t,\Delta x, z_*, p)\to
(\lambda^\nu \Delta t,\lambda \Delta x,\lambda^{1-\frac{\theta}{d-1}} z_*, \lambda^{d-1-\t}p).
\ee
Thus again the ratio $\frac{\Delta t}{(\Delta x)^\nu}$ is universal. Combining with numerical studies, we eventually find that this ratio has an upper bound as in (\ref{boundf}). Therefore the conclusion of the existence of extremal surfaces for the hyperscaling violation geometry is the same as that for the Lifshitz geometry.

\section{Conclusions}

In this paper we studied computations of holographic entanglement entropy and mutual information when we take a non-canonical time slice (refer to the right picture in Fig.\ref{fig:timeslice}). In the first half of this paper we studied them in CFTs at zero and finite temperature by using the AdS/CFT. We find that the mutual information $I(A,B)$ gets divergent as the separation of one end point of $A$ and that of $B$ becomes light-like. From the viewpoint of the deformed time slice which includes both $A$ and $B$ (refer to the left picture in Fig.\ref{fig:timeslice}), this light-like limit corresponds to the limit where the region $A$ and $B$ get causally touched with each other. Moreover, we confirmed that the same behavior in the light-like limit occurs in finite temperature holographic CFTs by studying extremal surfaces in AdS black branes.

In the latter half of this paper, we performed a similar analysis for gravity duals of non-relativistic scale invariant theories (Lifshitz geometries). Such a theory is characterized by the dynamical exponent $\nu$. In this case, the holographic entanglement entropy for a single interval or strip shows non-trivial behavior under a Lorentz boosting.
Note that for a pure AdS, the boosting a single subsystem is trivial as the theory is Lorentz invariant. By studying the extremal surfaces in Lifshitz geometries, we find that the extremal surface exists only if the time-like width $\Delta t$ and space-like width $\Delta x$ of the subsystem satisfy $\frac{\Delta t}{(\Delta x)^\nu}\leq f^{max}$. Here $f^{max}$ is a certain $O(1)$ constant, which depends on the dynamical exponent $\nu$ and the dimension $d$. The presence of this bound for the choices of subsystems in holographic entanglement entropy tells us that a real space decomposition of Hilbert space is not possible on generic time slices except the canonical ones. This may also suggest that we cannot define a quantum state in a standard way by taking any non-canonical time slice. Moreover we found that the same result is obtained for more general geometries so called the hyperscaling violation. It would be an interesting future problem to analyze the non-canonical time slice from the non-relativistic field theory viewpoint. At the same time, another interesting question is computations of boosted mutual information for more generic shapes of subsystems such as round balls \cite{Shiba,Cardy}, where quantum corrections in gravity will play an important role \cite{He,FLM,Fa} in addition to the classical contributions \cite{HiTa}.

Finally it is worth mentioning possible applications of our analysis. Our results of non-relativistic scale invariant theories show that if an observer is Lorentz boosted, 
then the observer will probe some sort of non-locality, which prevents us from a real space decomposition of the Hilbert space. This suggest that if we do an experiment of a condensed matter system at a non-relativistic critical point from a boosted observer, we may encounter interesting non-local effects. This deserves a future study.

\section*{ Acknowledgements}
We are grateful to Pawel Caputa, Naotaka Kubo, Tomonori Ugajin and Kento Watanabe for useful discussions.  TT is supported by the Simons Foundation through the ``It from Qubit'' collaboration, JSPS Grant-in-Aid for Scientific Research (A) No.16H02182 and World Premier International Research Center Initiative (WPI Initiative) from the Japan Ministry of Education, Culture, Sports, Science and Technology (MEXT).


\end{document}